\documentstyle[11pt,rotate,cite,epsfig]{article}
\begin{document}

\title{\Large\bf A proposal on the search for the hybrid with 
$I^G(J^{PC})=1^-(1^{-+})$ in the process 
$J/\psi\rightarrow\rho\omega\pi\pi$ at upgraded BEPC/BES \footnote{The 
project is supported by the National Natural Science Foundation of
China under Grant No. 19991487, and Grant No. LWTZ-1298 of the Chinese 
Academy of Sciences} } \vspace{1cm}

\author{ \small De-Min Li$^a$\footnote{E-mail: lidm@hptc5.ihep.ac.cn}, 
~~Hong Yu$^{a,b}$~ and~ Qi-Xing Shen$^{a,b}$\\ 
\small$^a$Institute of High Energy Physics, Chinese Academy of Sciences, \\
\small P.O.Box $918~(4)$, Beijing $100039$, China\footnote{Mailing address}\\
\small$^b$Institute of Theoretical Physics, Chinese Academy of Sciences, 
Bejing 100080, China } \maketitle
\vspace*{0.3cm}

\begin{abstract}

The moment expressions for the boson resonances $X$ 
with spin-parity $J_X^{P_XC}=0^{++}$, $1^{-+}$, $1^{++}$, and $2^{++}$ 
possibly produced in 
the process $J/\psi\rightarrow\rho X$, $X\rightarrow b_1(1235)\pi$, 
 $b_1\rightarrow \omega \pi$ are given in terms of the generalized moment 
analysis method. The resonance with $J_X^{P_XC}=1^{-+}$ can be 
distinguished from other resonances by means of these moments except for 
some rather special cases. The suggestion that the search for the hybrid with
$I^G(J^{PC})=1^-(1^{-+})$ can be performed in the decay channel
$J/\psi\rightarrow\rho\omega\pi\pi$ at upgraded BEPC/BES is presented.  

\end{abstract}

\vspace{1.5cm}



\newpage

\baselineskip 24pt

\section*{ I. Introduction}
\indent

Apart from the ordinary $q\bar{q}$ mesons, the new hadronic states such 
as glueballs ($gg/ggg$), hybrids ($q\bar{q}g)$ and four-quark states 
($qq\bar{q}\bar{q}$) also exist according to the predictions of QCD. 
Discovery and confirmation of any one of these new hadronic states would 
be the strong support to the QCD theory. Therefore, the search for and 
identifying these new hadronic states is a very excited and attractive 
research subject both theoretically and experimentally.

These new hadronic states can have the same quantum number $J^{PC}$ as the 
ordinary $q\bar{q}$ mesons, what's more, they can also have the exotic 
quantum 
number $J^{PC}$ which are not allowed in the quark model such as $1^{-+}$,
and thus they can not mix with the ordinary mesons. Experimentally, GAMS 
collaboration\cite{1}, E179 collaboration at KEK\cite{2}, VES 
group\cite{3}, E852 collaboration at BNL\cite{4} and Crystal Barrel\cite{5} 
all claimed that the evidence 
for the exotic state with $J^{PC}=1^{-+}$ was observed. The observed 
$\rho\pi$, $\eta\pi$ and $\eta^\prime\pi$ couplings of this state 
qualitatively support the hypothesis that it is a hybrid meson, although 
other interpretations cannot be eliminated\cite{6}. 

In terms of the predictions of the lattice QCD, the lowest lying glueball 
with $J^{PC}=1^{-+}$ has a higher mass than $J/\psi$\cite{7}. Bag model 
calculations\cite{8} 
predict that the lowest lying $qq\bar{q}\bar{q}$ states do not carry exotic 
quantum numbers and form nonets carrying the same quantum numbers as 
$q\bar{q}$ nonets, and that most $qq\bar{q}\bar{q}$ states can fall apart 
into 
two mesons and thus have a decay width in the order of their mass, which 
leads to that most $qq\bar{q}\bar{q}$ states are expected to be essentially 
unconfined and will not be observed as resonance peaks with reasonably narrow 
widths. Therefore, the search for the gluballs and four-quark 
states with $J^{PC}=1^{-+}$ at BEPC/BES could be disappointing. However, 
lattice QCD predicts that the mass of the hybrid with $J^{PC}=1^{-+}$ is 
$1.2\sim2.5$ GeV\cite{8}. In addition, the naive estimate of pQCD 
predicts 
that the $J/\psi$ hadronic decay processes are favorable to the production 
of hybrids. So, if the hybrids exist, the search for hybrid with 
$J^{PC}=1^{-+}$ at BEPC/BES should be fairly hopeful.  

H. Yu and Q.X. Shen\cite{9} have already discussed the possibility of 
the search for the hybrid with $J^{PC}=1^{-+}$ in the process 
$J/\psi\rightarrow\rho X$, 
$X\rightarrow\eta\pi~(\eta^\prime\pi,~\rho\pi)$. For the 
decay modes of the hybrid with $I^G(J^{PC})=1^-(1^{-+})$, according to 
the symmetrization selection rule\cite{10,11}, the 
$\eta\pi$, $\eta^\prime\pi$ 
modes are strongly suppressed (A possible mechanism to explain why 
the $1^{-+}$ state was observed in the above suppressed decay channels is 
planned 
for separated publication). The $\rho\pi$ mode is allowed, but this is 
a P-wave mode and thus the $\rho\pi$ mode should not be a dominant decay 
mode. The dominant decay mode should be the $b_1(1235)\pi$\cite{11}.
Therefore, the probability of discovering the hybrid with $J^{PC}=1^{-+}$ in 
the process $J/\psi\rightarrow\rho X$, $X\rightarrow b_1\pi$, in principle, 
should be higher than that in the process $J/\psi\rightarrow\rho X$, 
$X\rightarrow\eta\pi~ (\eta^\prime\pi,~\rho\pi)$.     
Also, since the dominant
decay mode of $b_1$ is $\omega\pi$, compared to the study on the two-step 
two-body decay process 
of $J/\psi$ in Ref. \cite{9,12}, the study on the three-step 
two-body decay process $J/\psi\rightarrow\rho
X$, $X\rightarrow b_1\pi$, $b_1\rightarrow\omega\pi$ perhaps could 
present more information to the experimentists. In this work, we shall 
consider the process $J/\psi\rightarrow\rho X$, $X\rightarrow b_1\pi$, 
$b_1\rightarrow\omega\pi$.

This work is organized as follows. In Sect. II, we give the moment
expressions for the resonances $X$ with the above spin-parity in the 
process $J/\psi\rightarrow\rho
X$, $X\rightarrow b_1\pi$, $b_1\rightarrow\omega\pi$ in terms of the
generalized moment analysis method\cite{9,12,13}.
, and in Sect. III, we discuss how to identify the resonances $X$ with 
different spin-parity. Our conclusion is reached in Sect. IV.

\section*{II. Moment analysis}
\indent

We consider the process 
\begin{equation}
e^+~+~e^-~\rightarrow J/\psi\rightarrow ~\rho~+~X, ~~X\rightarrow 
~b_1~+~\pi, ~~b_1\rightarrow\omega~+~\pi.
\end{equation}
The S matrix element of the process (1) can be 
written as 
\begin{equation}
\langle\rho_{\lambda_\rho}\omega_{\lambda_\omega}\pi\pi|S-1|
e^+_re^-_{r^\prime}\rangle\propto
\langle\psi_{\lambda_J}|T|e^+_re^-_{r^\prime}\rangle
\langle\rho_{\lambda_\rho}X_{\lambda_X}|T_1|\psi_{\lambda_J}\rangle
\langle b_{1\lambda_{b_1}}\pi|T_2|X_{\lambda_X}\rangle
\langle\omega_{\lambda_\omega}\pi|T_3|b_{1\lambda_{b_1}}\rangle,
\end{equation}
where 
\begin{equation}
\langle\psi_{\lambda_J}|T|e^+_re^-_{r^\prime}\rangle
\propto e^{\lambda_J\ast}_{\mu}(\vec{p}_J)\bar{v}_r(\vec{p}_+)\gamma^{\mu}
u_{r^\prime}(\vec{p}_-);
\end{equation}
\begin{equation}
\langle\rho_{\lambda_\rho}X_{\lambda_X}|T_1|\psi_{\lambda_J}\rangle
\propto 
A^{J_X}_{\lambda_\rho,\lambda_X}
D^{1\ast}_{\lambda_J,\lambda_\rho-\lambda_X}(0,\theta_\rho,0);
\end{equation}
\begin{equation}
\langle b_{1\lambda_{b_1}}\pi|T_2|X_{\lambda_X}\rangle
\propto B^{J_X}_{\lambda_{b_1}}D^{J_X\ast}_{\lambda_X,\lambda_{b_1}}(\phi_1,
\theta_1,-\phi_1);
\end{equation}
\begin{equation}
\langle\omega_{\lambda_\omega}\pi|T_3|b_{1\lambda_{b_1}}\rangle
\propto C_{\lambda_\omega}D^{1\ast}_{\lambda_{b_1},\lambda_\omega}
(\phi_2,\theta_2, -\phi_2);
\end{equation}
And $\lambda_J$, $\lambda_\rho$, $\lambda_X$, 
$\lambda_{b_1}$ and 
$\lambda_\omega$ are the helicities of $J/\psi$, $\rho$, $X$, $b_1$ and 
$\omega$, respectively; $r$ and $r^\prime$ are the polarization indexes 
of the positron and electron, respectively; 
$\vec{p}_J$, $\vec{p}_+$, $\vec{p}_-$ are the 
momenta of $J/\psi$, positron and electron in 
the c.m. system of $e^+e^-$, respectively; 
$A^{J_X}_{\lambda_\rho,\lambda_X}$, 
$B^{J_X}_{\lambda_{b_1}}$ and $C_{\lambda_\omega}$ are the helicity 
amplitudes 
of the processed $J/\psi\rightarrow\rho X$, $X\rightarrow b_1\pi$ and 
$b_1\rightarrow \omega\pi$, respectively; $\theta_\rho$ is the polar angle 
in the c.m. system of $e^+e^-$ in which $z$ axis is chosen to be along 
the direction of the 
incident positron and the vector meson $\rho$ lies in $x-z$ 
plane; $(\theta_1,\phi_1)$ describes the direction of the momentum of $b_1$ 
in the rest frame of $X$ where the $z_1$ axis is chosen to be along the 
direction of the momentum of $X$ in the c.m. system of $e^+e^-$; 
Similarly, $(\theta_2,\phi_2)$ described the direction of the momentum of 
the vector mesons $\omega$ in the rest frame of $b_1$ where the $z_2$ 
axis is along the momentum of $b_1$ in the rest frame of $X$; The 
function $D^J_{m,n}$ is the $(2J+1)$-dimensional representation of the 
rotation group. Owing to the parity conservation for the process (1), 
these helicity amplitudes satisfy the following symmetry 
relations\cite{14}: 
\begin{eqnarray}
&&A^{J_X}_{-\lambda_\rho,-\lambda_X}=
P_X(-1)^{J_X}A^{J_X}_{\lambda_\rho,\lambda_X},
\nonumber\\
&&B^{J_X}_{-\lambda_{b_1}}=P_X(-1)^{J_X}B^{J_X}_{\lambda_{b_1}},\nonumber\\
&&C_{-\lambda_\omega}=C_{\lambda_\omega},
\end{eqnarray}
where $P_X$ is the parity of $X$.
 
The angular distribution for the process (1) is 
\begin{eqnarray}
&&W(\theta_\rho,\theta_1,\phi_1,\theta_2,\phi_2)\propto\nonumber\\
&&\sum_{\lambda_J,\lambda^\prime_J}\sum_{\lambda_X,\lambda^\prime_X}
\sum_{\lambda_{b_1},\lambda^\prime_{b_1}}\sum_{\lambda_\rho,\lambda_\omega}
I_{\lambda_J,\lambda^\prime_J}A^{J_X}_{\lambda_\rho,\lambda_X}
A^{J_X\ast}_{\lambda_\rho,\lambda^\prime_X}
B^{J_X}_{\lambda_{b_1}}B^{J_X\ast}_{\lambda^\prime_{b_1}}
C_{\lambda_\omega}C^\ast_{\lambda_\omega}
\nonumber\\
&&\times D^{1\ast}_{\lambda_J,\lambda_\rho-\lambda_X}(0,\theta_\rho,0)
D^{1}_{\lambda^\prime_J,\lambda_\rho-\lambda^\prime_X}(0,\theta_\rho.0)
\nonumber\\
&&\times D^{J_X\ast}_{\lambda_X,\lambda_{b_1}}(\phi_1,\theta_1,-\phi_1)
D^{J_X}_{\lambda^\prime_X,\lambda^\prime_{b_1}}(\phi_1,\theta_1,-\phi_1)
\nonumber\\
&&\times D^{1\ast}_{\lambda_{b_1},\lambda_\omega}(\phi_2,\theta_2,-\phi_2)
D^{1}_{\lambda^\prime_{b_1},\lambda_\omega}(\phi_2,\theta_2,-\phi_2),
\end{eqnarray}
where the density matrix elements $I_{\lambda_J,\lambda^\prime_J}$ is
\begin{eqnarray}
I_{\lambda_J,\lambda^\prime_J}\equiv \frac{1}{4}\sum_{r,r^\prime}
\langle\psi_{\lambda_J}|T|e^+_re^-_{r^\prime}\rangle
{\langle\psi_{\lambda^\prime_J}|T|e^+_re^-_{r^\prime}\rangle}^\ast
\propto
 2|\vec{p}_+|^2\delta_{\lambda_J,\lambda^\prime_J}\delta_{\lambda_J,\pm1}.
\end{eqnarray}
The moments for the process (1) can be defined by
\begin{eqnarray}
&&M(j,L,M,\ell,m)=
\nonumber\\
&&\int d\theta_\rho\sin\theta_\rho d\theta_1\sin\theta_1 d\phi_1 
d\theta_2\sin\theta_2 d\phi_2 W(\theta_\rho,\theta_1,\phi_1,\theta_2,\phi_2)
\nonumber\\
&&\times D^j_{0,-M}(0,\theta_\rho,0)D^L_{M,m}(\phi_1,\theta_1,-\phi_1)
D^\ell_{m,0}(\phi_2,\theta_2,-\phi_2). 
\end{eqnarray}
Eq. (10) can be reduced to 
\begin{eqnarray}
&&M(j,L,M,\ell,m)\propto
\nonumber\\
&&\sum_{\lambda_J=\pm 1}\sum_{\lambda_X,\lambda^\prime_X}
\sum_{\lambda_{b_1},\lambda^\prime_{b_1}}\sum_{\lambda_\rho,\lambda_\omega}
A^{J_X}_{\lambda_\rho,\lambda_X}A^{J_X\ast}_{\lambda_\rho,\lambda^\prime_X}
B^{J_X}_{\lambda_{b_1}}B^{J_X\ast}_{\lambda^\prime_{b_1}}C_{\lambda_\omega}
C^\ast_{\lambda_\omega}
\nonumber\\
&&\times\langle 1\lambda_Jj0|1\lambda_J\rangle
\langle 1(\lambda_\rho-\lambda^\prime_X)j(-M)|
1(\lambda_\rho-\lambda_X)\rangle
\nonumber\\
&&\times\langle J_X\lambda^\prime_XLM|J_X\lambda_X\rangle
\langle J_X\lambda^\prime_{b_1}Lm|J_X\lambda_{b_1}\rangle 
\nonumber\\
&&\times\langle 1\lambda^\prime_{b_1}\ell m|1\lambda_{b_1}\rangle
\langle 1\lambda_\omega \ell 0|1\lambda_\omega\rangle,
\end{eqnarray}
where $\langle j_1m_1j_2m_2|j_3m_3\rangle$ is Clebsch-Gordan coefficients.

In the process $X\rightarrow b_1\pi$, if we restrict $\ell_f\leq 1$, where 
$\ell_f$ is the relative orbital angular momentum between $b_1$ and $\pi$, 
 the quantum number $I^G(J_X^{P_XC})$ of $X$ allowed by the 
parity-isospin conservation law in 
the process (1) are $1^-(1^{-+})$, $1^-(0^{++})$, 
$1^-(1^{++})$, and $1^-(2^{++})$. For the resonances $X$ with 
$J^{P_XC}_X=0^{++},~1^{-+},~1^{++}$, and $2^{++}$, the nonzero moment 
expressions derived from Eq. (11) are shown in Appendix A, B, C, and D.

There are four, twenty-one, sixteen, and thirty-one nonzero moment 
expressions for $J_X^{P_XC}=0^{++}$, $1^{-+}$, $1^{++}$, and $2^{++}$, 
respectively. 
In the following section, we shall discuss how to identify the $X$ with 
the above $J_X^{P_XC}$.
 
\section*{III. Discussion}
\indent

Since the helicity amplitudes $|C_0|^2$ and $|C_1|^2$ are independent of 
the spin-parity of the resonance $X$,
we find that if 
$|C_0|^2\not=|C_1|^2$ the moment expressions have the following 
characteristics:  
For $J_X^{P_XC}=0^{++}$, the moments is always equal to zero in 
the case $L>0$ or $M>0$ or $m>0$; For $J_X^{P_XC}=1^{++}$, the nonzero 
moments with $L=0,1,2$, $M=0,1,2$ and $m=0,2$ exist but the moments are 
zero in the case $m=1$; For $J_X^{P_XC}=1^{-+}$, the nonzero moments with
$L=0,1,2$, $M=0,1,2$ and $m=0,2$ exist, the nonzero moments with $m=1$
also exist;  
For $J_X^{P_XC}=2^{++}$, apart from the nonzero moments with
$L=0,1,2$, $M=0,1,2$ and $m=0,1,2$, the nonzero moments with
$L=3,4$ exist. Therefore, from these characteristics, we can easily identify 
the resonances $X$ with $J_X^{P_XC}$= $0^{++}$, $1^{-+}$, $1^{++}$ and 
$2^{++}$ experimentally. 

However, if $|C_0|^2=|C_1|^2$, some of the preceding characteristics 
disappear, which leads to that the situations in the case 
$|C_0|^2=|C_1|^2$ are more complex 
than those in the case $|C_0|^2\not=|C_1|^2$. We will turn to the 
special case $|C_0|^2=|C_1|^2$ below.
 
(A) $|C_0|^2=|C_1|^2$, $|B^1_0|^2\not=|B^1_1|^2$ and 
$3|B^2_0|^2\not=4|B^2_1|^2$

In this case, only for $J_X^{P_XC}=2^{++}$, there are four nonzero 
moments with $L=4$, so the resonance with $J_X^{P_XC}=2^{++}$ can be 
distinguished from other resonances. Then, for $J_X^{P_XC}=0^{++}$, 
there are only two nonzero moments with $L=0$, and for 
$J_X=1$, there are four nonzero moments with $L=2$, in addition to two 
nonzero moments with $L=0$, hence the resonance with $J_X=0$ can also be 
distinguished from that with $J_X=1$. Finally, to distinguish the 
resonance with $J_X^{P_XC}=1^{-+}$ from that with $J_X^{P_X}=1^{++}$, we 
consider the following moment expression
$H\equiv \frac{1}{8}M(00000)-\frac{5}{4}M(02000)-\frac{5}{4}M(20000)+
\frac{25}{2}M(22000)$ and find that the $H$ satisfies
\begin{equation}
H\propto\left\{\begin{array}{cc}
0,& (J_X^{P_XC}=1^{++}),\\
-\frac{27}{4}(|A^1_{00}|^2|B^1_{0}|^2-2|A^1_{11}|^2|B^1_{0}|^2-
2|A^1_{00}|^2|B^1_{1}|^2)|C_1|^2, & 
(J_X^{P_XC}=1^{-+}).
 \end{array}\right.
\end{equation}
Using Eq. (12), we can still distinguish the resonance $X$ with 
$J_X^{P_XC}=1^{-+}$ from that with $J_X^{P_XC}=1^{++}$.

(B) $|C_0|^2=|C_1|^2$, $|B^1_0|^2\not=|B^1_1|^2$ and 
$3|B^2_0|^2=4|B^2_1|^2$

In this case, compared to the case (A), 
the numbers of the nonzero moments for $J_X^{P_XC}=0^{++}$, $1^{-+}$ and 
$1^{++}$ remain unchange, but for $J_X^{P_XC}=2^{++}$, 
the moments with $L=4$ disappear, There are still only two nonzero moments 
with $L=0$ for $J_X^{P_XC}=0^{++}$ and six nonzero moments with $L=0,~2$ 
not only for $J_X=1$ but also for $J_X=2$. In this case, owing to Eq. 
(12) remains unchange and  
\begin{equation}
H\propto-\frac{15}{2}|A^2_{00}|^2|B^2_{1}|^2|C_1|^2<0~~(J_X^{P_XC}=2^{++}),
\end{equation}
the crucial point
is to distinguish the resonance with $J_X^{P_XC}=1^{-+}$ from that with
$J_X^{P_XC}=2^{++}$. We also find
$H_1\equiv \frac{1}{4}M(02000)-\frac{5}{2}M(22000)$ satisfies
\begin{equation}
H_1\propto\left\{\begin{array}{cc}
3(|A^2_{00}|^2+|A^2_{11}|^2)|B^2_{1}|^2|C_{1}|^2>0, &
(J_X^{P_XC}=2^{++}),\\
\frac{9}{5}|A^1_{11}|^2|B^1_1|^2|C_1|^2 >0, &
(J_X^{P_XC}=1^{++}),\\ 
-\frac{9}{5}(|A^1_{00}|^2-|A^1_{11}|^2)(|B_1^1|^2-|B^1_0|^2)|C_1|^2, &
(J_X^{P_XC}=1^{-+}).
\end{array}
\right.
\end{equation}
So, if it is determined 
experimentally that $H>0$ or $H_1\leq 0$, from Eq.(12)$\sim$(14), the 
$J_X^{P_XC}$ of $X$ must be $1^{-+}$. However, if $H<0$ or $H_1>0$,   
we can not distinguish the resonance with $J_X^{P_XC}=2^{++}$ from that 
with $J_X^{P_XC}=1^{-+}$.

(C) $|C_0|^2=|C_1|^2$ and $|B^1_0|^2=|B^1_1|^2$
 
In this case, there are only two nonzero moments with $L=M=\ell=m=0$ both 
for $J_X^{P_XC}=0^{++}$ and $J_X^{P_XC}=1^{-+}$, there are two nonzero 
moments with $L=0$ and four nonzero moments with $L=2$ for 
$J_X^{P_XC}=1^{++}$, and there are two nonzero moments with $L=0$ and at 
least four nonzero moments with $L=2$ for $J_X^{P_XC}=2^{++}$. Therefore, 
the resonances with $J_X^{P_XC}=1^{++}$ and $2^{++}$ can be distinguished 
from the resonance with $J_X^{P_XC}=0^{++}$ ( or $1^{-+}$ ). But it is 
almost impossible to distinguish 
the resonance with $J_X^{P_XC}=1^{-+}$ from that with $J_X^{P_XC}=0^{++}$ 
except in the radiative $J/\psi$
decay process. Because for the radiative $J/\psi$
decay process $e^++e^-\rightarrow J/\psi\rightarrow \gamma X$,
$X\rightarrow b_1\pi$, $b_1\rightarrow \omega\pi$, 
$A^0_{00}=A^0_{01}=A^1_{00}=A^1_{01}$,
we find
\begin{equation}
M(00000)-10M(20000)\propto\left\{\begin{array}{cc}
0, & (J_X^{P_XC}=0^{++}),\\
108|A^1_{11}|^2|B^1_1|^2|C_1|^2, & (J_X^{P_XC}=1^{-+}).
\end{array}\right.
\end{equation} 
Obviously, using Eq. (15) we can distinguish the 
$0^{++}$ state from the $1^{-+}$ state in the radiative $J/\psi$ decay 
process.

From the above discussions, we get 
that if $|C_0|^2\not=|C_1|^2$ we can 
easily identify the resonances $X$ with $J_X^{P_XC}$= $0^{++}$, $1^{-+}$, 
$1^{++}$ and $2^{++}$, but if 
$|C_0|^2=|C_1|^2$ and $|B^1_0|^2=|B^1_1|^2$ ( or $|C_0|^2=|C_1|^2$ and 
$3|B^2_0|^2=4|B^2_1|^2$ ), the identification of the 
resonances $X$ with $J_X^{P_XC}$=$1^{-+}$ and $0^{++}$ ( or $2^{++}$ ) is 
very difficult. However, we also want to note the 
following  
two points: 1) Since the ratio of the helicity amplitudes for the process
$b_1(1235)\rightarrow\omega\pi$, $|C_0|$ and $|C_1|$, can be measured
experimentally in other process such as $J/\psi\rightarrow b_1\pi$,
$b_1\rightarrow\omega\pi$. 
The measurement of the ratio of $|C_0|$ and
$|C_1|$ can be first performed in order to confirm whether $|C_0|^2$ is 
equal to $|C_1|^2$ or not;
2) Even though $|C_0|^2=|C_1|^2$, 
one could expect that the probability of the simultaneous appearance
of $|C_0|^2=|C_1|^2$ and $|B^1_0|^2=|B^1_1|^2$ ( or $|C_0|^2=|C_1|^2$ and 
$3|B^2_0|^2=4|B^2_1|^2$ ) would be fairly small.

It is worth pointing out that the above moment expressions and 
the discussions are also valid for the process $J/\psi\rightarrow\gamma X$, 
 $X\rightarrow b_1\pi$, $b_1\rightarrow\omega\pi$ provided 
$A^0_{00}=A^0_{01}=A^1_{00}=A^1_{01}=A^2_{00}=A^2_{01}=0$.

\section*{IV. Conclusion} 
\indent

The twenty-one nonzero moment expressions for $J_X^{P_XC}=1^{-+}$ show the 
possibility of the resonance $X$ with $J_X^{P_XC}=1^{-+}$ produced in the 
process (1) exists. At the same time, we can easily distinguish it from 
other resonances except for some rather special cases. Therefore, generally 
speaking, if the 50 million $J/\psi$ events in the upgraded BEPC/BES are 
obtained, the search for the hybrid with $J^{PC}=1^{-+}$ in 
the process $J/\psi\rightarrow\rho X$, $X\rightarrow b_1\pi$, 
$b_1(1235)\rightarrow\omega\pi$ is feasible. 

\section*{ Appendix A: The nonzero moments for $J_X^{P_XC}=0^{++}$}
\begin{eqnarray} 
&&M(00000)\propto
2(|A^0_{00}|^2+2|A^0_{10}|^2)|B^0_0|^2(|C_0|^2+2|C_1|^2),
\nonumber\\
&&M(00020)\propto
\frac{4}{5}(|A^0_{00}|^2+2|A^0_{10}|^2)|B^0_0|^2(|C_0|^2-|C_1|^2), 
\nonumber\\
&&M(20000)\propto
-\frac{2}{5}(|A^0_{00}|^2-|A^0_{10}|^2)|B^0_0|^2(|C_0|^2+2|C_1|^2),
\nonumber\\
&&M(20020)\propto
-\frac{4}{25}(|A^0_{00}|^2-|A^0_{10}|^2)|B^0_0|^2(|C_0|^2-|C_1|^2).
\nonumber
\end{eqnarray}   

\section*{ Appendix B: The nonzero moments for $J_X^{P_XC}=1^{-+}$}

\begin{eqnarray}
&&M(00000)\propto
2(|A^1_{00}|^2+2|A^1_{01}|^2+2|A^1_{10}|^2+2|A^1_{11}|^2)
(|B^1_0|^2+2|B^1_1|^2)(|C_0|^2+2|C_1|^2),
\nonumber\\
&&M(00020)\propto
\frac{4}{5}(|A^1_{00}|^2+2|A^1_{01}|^2+2|A^1_{10}|^2+2|A^1_{11}|^2)  
(|B^1_0|^2-|B^1_1|^2)(|C_0|^2-|C_1|^2),
\nonumber\\
&&M(02000)\propto
\frac{4}{5}(|A^1_{00}|^2-|A^1_{01}|^2+2|A^1_{10}|^2-|A^1_{11}|^2)
(|B^1_0|^2-|B^1_1|^2)(|C_0|^2+2|C_1|^2),
\nonumber\\
&&M(02020)\propto
\frac{4}{25}(|A^1_{00}|^2-|A^1_{01}|^2+2|A^1_{10}|^2-|A^1_{11}|^2)
(2|B^1_0|^2+|B^1_1|^2)(|C_0|^2-|C_1|^2),
\nonumber\\
&&M(02021)\propto
\frac{12}{25}(|A^1_{00}|^2-|A^1_{01}|^2+2|A^1_{10}|^2-|A^1_{11}|^2)
Re(B^1_1B^{1\ast}_0)(|C_0|^2-|C_1|^2),
\nonumber\\
&&M(02022)\propto
\frac{12}{25}(|A^1_{00}|^2-|A^1_{01}|^2+2|A^1_{10}|^2-|A^1_{11}|^2)
|B^1_1|^2(|C_0|^2-|C_1|^2),
\nonumber\\
&&M(20000)\propto
-\frac{2}{5}(|A^1_{00}|^2-|A^1_{01}|^2-|A^1_{10}|^2+2|A^1_{11}|^2)
(|B^1_0|^2+2|B^1_1|^2)(|C_0|^2+2|C_1|^2),
\nonumber\\
&&M(20020)\propto
-\frac{4}{25}(|A^1_{00}|^2-|A^1_{01}|^2-|A^1_{10}|^2+2|A^1_{11}|^2)
(|B^1_0|^2-|B^1_1|^2)(|C_0|^2-|C_1|^2),
\nonumber\\
&&M(21121)\propto
\frac{6}{25}Im(A^1_{01}A^{1\ast}_{00}+A^1_{10}A^{1\ast}_{11})
Im(B^1_1B^{1\ast}_0)(|C_0|^2-|C_1|^2),
\nonumber\\
&&M(22000)\propto
-\frac{2}{25}(2|A^1_{00}|^2+|A^1_{01}|^2-2|A^1_{10}|^2-2|A^1_{11}|^2)
(|B^1_0|^2-|B^1_1|^2)(|C_0|^2+2|C_1|^2),
\nonumber\\
&&M(22020)\propto
-\frac{2}{125}(2|A^1_{00}|^2+|A^1_{01}|^2-2|A^1_{10}|^2-2|A^1_{11}|^2)
(2|B^1_0|^2+|B^1_1|^2)(|C_0|^2-|C_1|^2),
\nonumber\\
&&M(22021)\propto
-\frac{6}{125}(2|A^1_{00}|^2+|A^1_{01}|^2-2|A^1_{10}|^2-2|A^1_{11}|^2)
Re(B^1_1B^{1\ast}_0)(|C_0|^2-|C_1|^2),
\nonumber\\
&&M(22022)\propto
-\frac{6}{125}(2|A^1_{00}|^2+|A^1_{01}|^2-2|A^1_{10}|^2-2|A^1_{11}|^2)
|B^1_1|^2(|C_0|^2-|C_1|^2),
\nonumber\\
&&M(22100)\propto
-\frac{6}{25}Re(A^1_{01}A^{1\ast}_{00}-A^1_{11}A^{1\ast}_{10})
(|B^1_0|^2-|B^1_1|^2)(|C_0|^2+2|C_1|^2),
\nonumber\\
&&M(22120)\propto
-\frac{6}{125}Re(A^1_{01}A^{1\ast}_{00}-A^1_{11}A^{1\ast}_{10})
(2|B^1_0|^2+|B^1_1|^2)(|C_0|^2-|C_1|^2),
\nonumber\\
&&M(22121)\propto
-\frac{18}{125}Re(A^1_{01}A^{1\ast}_{00}-A^1_{11}A^{1\ast}_{10})
Re(B^1_1B^{1\ast}_0)(|C_0|^2-|C_1|^2),
\nonumber\\
&&M(22122)\propto
-\frac{18}{125}Re(A^1_{01}A^{1\ast}_{00}-A^1_{11}A^{1\ast}_{10})
|B^1_1|^2(|C_0|^2-|C_1|^2),
\nonumber\\
&&M(22200)\propto
-\frac{6}{25}|A^1_{01}|^2
(|B^1_0|^2-|B^1_1|^2)(|C_0|^2+2|C_1|^2),
\nonumber\\
&&M(22220)\propto
-\frac{6}{125}|A^1_{01}|^2
(2|B^1_0|^2+|B^1_1|^2)(|C_0|^2-|C_1|^2),
\nonumber\\
&&M(22221)\propto
-\frac{18}{125}|A^1_{01}|^2
Re(B^1_1B^{1\ast}_0)(|C_0|^2-|C_1|^2),
\nonumber\\
&&M(22222)\propto
-\frac{18}{125}|A^1_{01}|^2
|B^1_1|^2(|C_0|^2-|C_1|^2).
\nonumber
\end{eqnarray}

\section*{ Appendix C: The nonzero moments for $J_X^{P_XC}=1^{++}$}
\begin{eqnarray}
&&M(00000)\propto
8(|A^1_{01}|^2+|A^1_{10}|^2+|A^1_{11}|^2)
|B^1_1|^2(|C_0|^2+2|C_1|^2),
\nonumber\\
&&M(00020)\propto
-\frac{8}{5}(|A^1_{01}|^2+|A^1_{10}|^2+|A^1_{11}|^2)
|B^1_1|^2(|C_0|^2-|C_1|^2),
\nonumber\\
&&M(02000)\propto
\frac{4}{5}(|A^1_{01}|^2-2|A^1_{10}|^2+|A^1_{11}|^2)
|B^1_1|^2(|C_0|^2+2|C_1|^2),
\nonumber\\
&&M(02020)\propto
-\frac{4}{25}(|A^1_{01}|^2-2|A^1_{10}|^2+|A^1_{11}|^2)
|B^1_1|^2(|C_0|^2-|C_1|^2),
\nonumber\\
&&M(02022)\propto
\frac{12}{25}(|A^1_{01}|^2-2|A^1_{10}|^2+|A^1_{11}|^2)
|B^1_1|^2(|C_0|^2-|C_1|^2),
\nonumber\\
&&M(20000)\propto
\frac{4}{5}(|A^1_{01}|^2+|A^1_{10}|^2-2|A^1_{11}|^2)
|B^1_1|^2(|C_0|^2+2|C_1|^2),
\nonumber\\
&&M(20020)\propto
-\frac{4}{25}(|A^1_{01}|^2+|A^1_{10}|^2-2|A^1_{11}|^2)
|B^1_1|^2(|C_0|^2-|C_1|^2),
\nonumber\\
&&M(22000)\propto
\frac{2}{25}(|A^1_{01}|^2-2|A^1_{10}|^2-2|A^1_{11}|^2)
|B^1_1|^2(|C_0|^2+2|C_1|^2),
\nonumber\\
&&M(22020)\propto
-\frac{2}{125}(|A^1_{01}|^2-2|A^1_{10}|^2-2|A^1_{11}|^2)
|B^1_1|^2(|C_0|^2-|C_1|^2),
\nonumber\\
&&M(22022)\propto
\frac{6}{125}(|A^1_{01}|^2-2|A^1_{10}|^2-2|A^1_{11}|^2)
|B^1_1|^2(|C_0|^2-|C_1|^2),
\nonumber\\
&&M(22100)\propto
-\frac{6}{25}Re(A^1_{11}A^{1\ast}_{10})
|B^1_1|^2(|C_0|^2+2|C_1|^2),
\nonumber\\
&&M(22120)\propto
\frac{6}{125}Re(A^1_{11}A^{1\ast}_{10})
|B^1_1|^2(|C_0|^2-|C_1|^2),
\nonumber\\
&&M(22122)\propto
-\frac{18}{125}Re(A^1_{11}A^{1\ast}_{10})
|B^1_1|^2(|C_0|^2-|C_1|^2),
\nonumber\\
&&M(22200)\propto
-\frac{6}{25}|A^1_{01}|^2
|B^1_1|^2(|C_0|^2+2|C_1|^2),
\nonumber\\
&&M(22220)\propto
\frac{6}{125}|A^1_{01}|^2
|B^1_1|^2(|C_0|^2-|C_1|^2),
\nonumber\\   
&&M(22222)\propto
-\frac{18}{125}|A^1_{01}|^2
|B^1_1|^2(|C_0|^2-|C_1|^2).
\nonumber
\end{eqnarray}

\section*{ Appendix D: The nonzero moments for $J_X^{P_XC}=2^{++}$}

\begin{eqnarray}
&&M(00000)\propto
2(|A^2_{00}|^2+2|A^2_{01}|^2+2|A^2_{10}|^2+2|A^2_{11}|^2+2|A^2_{12}|^2)
(|B^2_0|^2+2|B^2_1|^2)(|C_0|^2+2|C_1|^2),
\nonumber\\
&&M(00020)\propto
\frac{4}{5}(|A^2_{00}|^2+2|A^2_{01}|^2+2|A^2_{10}|^2+2|A^2_{11}|^2+ 
2|A^2_{12}|^2)
(|B^2_0|^2-|B^2_1|^2)(|C_0|^2-|C_1|^2),
\nonumber\\
&&M(02000)\propto   
\frac{4}{7}(|A^2_{00}|^2+|A^2_{01}|^2+2|A^2_{10}|^2+|A^2_{11}|^2
-2|A^2_{12}|^2)
(|B^2_0|^2+|B^2_1|^2)(|C_0|^2+2|C_1|^2),
\nonumber\\
&&M(02020)\propto
\frac{4}{35}
(|A^2_{00}|^2+|A^2_{01}|^2+2|A^2_{10}|^2+|A^2_{11}|^2-2|A^2_{12}|^2)
(2|B^2_0|^2-|B^2_1|^2)(|C_0|^2-|C_1|^2),
\nonumber\\
&&M(02021)\propto
\frac{4\sqrt{3}}{35}
(|A^2_{00}|^2+|A^2_{01}|^2+2|A^2_{10}|^2+|A^2_{11}|^2-2|A^2_{12}|^2)
Re(B^2_1B^{2\ast}_0)(|C_0|^2-|C_1|^2),
\nonumber\\
&&M(02022)\propto
\frac{12}{35}(|A^2_{00}|^2+|A^2_{01}|^2+2|A^2_{10}|^2
+|A^2_{11}|^2-2|A^2_{12}|^2)
|B^2_1|^2(|C_0|^2-|C_1|^2),
\nonumber\\
&&M(04000)\propto
\frac{4}{63}(3|A^2_{00}|^2-4|A^2_{01}|^2+6|A^2_{10}|^2
-4|A^2_{11}|^2+|A^2_{12}|^2)
(3|B^2_0|^2-4|B^2_1|^2)(|C_0|^2+2|C_1|^2),
\nonumber\\
&&M(04020)\propto
\frac{8}{315}(3|A^2_{00}|^2-4|A^2_{01}|^2+6|A^2_{10}|^2
-4|A^2_{11}|^2+|A^2_{12}|^2)
(3|B^2_0|^2+2|B^2_1|^2)(|C_0|^2-|C_1|^2),
\nonumber\\
&&M(04021)\propto
\frac{4\sqrt{10}}{105}
(3|A^2_{00}|^2-4|A^2_{10}|^2+6|A^2_{10}|^2-4|A^2_{11}|^2+|A^2_{12}|^2)
Re(B^2_1B^{2\ast}_0)(|C_0|^2-|C_1|^2),
\nonumber\\
&&M(04022)\propto
\frac{8\sqrt{15}}{315}
(3|A^2_{00}|^2-4|A^2_{10}|^2+6|A^2_{10}|^2-4|A^2_{11}|^2+|A^2_{12}|^2)
|B^2_1|^2(|C_0|^2-|C_1|^2),
\nonumber\\
&&M(20000)\propto
-\frac{2}{5}(|A^2_{00}|^2-|A^2_{01}|^2
-|A^2_{10}|^2+2|A^2_{11}|^2-|A^2_{12}|^2)
(|B^2_0|^2+2|B^2_1|^2)(|C_0|^2+2|C_1|^2),
\nonumber\\
&&M(20020)\propto
-\frac{4}{25}
(|A^2_{00}|^2-|A^2_{01}|^2-|A^2_{10}|^2+2|A^2_{11}|^2-|A^2_{12}|^2)
(|B^2_0|^2-|B^2_1|^2)(|C_0|^2-|C_1|^2),
\nonumber\\
&&M(21121)\propto
\frac{2}{25}[3Im(A^2_{01}A^{2\ast}_{00}+A^2_{10}A^{2\ast}_{11})+
\sqrt{6}Im(A^2_{12}A^{2\ast}_{11})]
Im(B^2_1B^{2\ast}_0)(|C_0|^2-|C_1|^2),
\nonumber\\
&&M(22000)\propto
-\frac{2}{35}(2|A^2_{00}|^2-|A^2_{01}|^2
-2|A^2_{10}|^2+2|A^2_{11}|^2+2|A^2_{12}|^2)
(|B^2_0|^2+|B^2_1|^2)(|C_0|^2+2|C_1|^2),
\nonumber\\
&&M(22020)\propto
-\frac{2}{175}
(2|A^2_{00}|^2-|A^2_{01}|^2-2|A^2_{10}|^2+2|A^2_{11}|^2+2|A^2_{12}|^2)
(2|B^2_0|^2-|B^2_1|^2)(|C_0|^2-|C_1|^2),
\nonumber\\   
&&M(22021)\propto
-\frac{2\sqrt{3}}{175}
(2|A^2_{00}|^2-|A^2_{01}|^2-2|A^2_{10}|^2+2|A^2_{11}|^2+2|A^2_{12}|^2)
Re(B^2_1B^{2\ast}_0)(|C_0|^2-|C_1|^2),
\nonumber\\
&&M(22022)\propto
-\frac{6}{175}(2|A^2_{00}|^2-|A^2_{01}|^2-2|A^2_{10}|^2+2|A^2_{11}|^2
+2|A^2_{12}|^2)
|B^2_1|^2(|C_0|^2-|C_1|^2),
\nonumber\\
&&M(22100)\propto
-\frac{\sqrt{2}}{35}
[\sqrt{6}Re(A^2_{01}A^{2\ast}_{00}-A^2_{11}A^{2\ast}_{10})
+6Re(A^2_{12}A^{2\ast}_{11})]
(|B^2_0|^2+|B^2_1|^2)(|C_0|^2+2|C_1|^2),
\nonumber\\
&&M(22120)\propto
-\frac{2}{175}
[\sqrt{3}Re(A^2_{01}A^{2\ast}_{00}-A^2_{11}A^{2\ast}_{10})
+3\sqrt{2}Re(A^2_{12}A^{2\ast}_{11})]
(2|B^2_0|^2-|B^2_1|^2)(|C_0|^2-|C_1|^2),
\nonumber\\
&&M(22121)\propto
-\frac{6}{175}[Re(A^2_{01}A^{2\ast}_{00}-A^2_{11}A^{2\ast}_{10})
+\sqrt{6}Re(A^2_{12}A^{2\ast}_{11})]
Re(B^2_1B^{2\ast}_0)(|C_0|^2-|C_1|^2),
\nonumber\\
&&M(22122)\propto
-\frac{3\sqrt{2}}{175}
[\sqrt{6}Re(A^2_{01}A^{2\ast}_{00}-A^2_{11}A^{2\ast}_{10})
+6Re(A^2_{12}A^{2\ast}_{11})]
|B^2_1|^2(|C_0|^2-|C_1|^2),
\nonumber\\
&&M(22200)\propto
-\frac{2}{35}
[3|A^2_{01}|^2+2\sqrt{6}Re(A^2_{12}A^{2\ast}_{10})]
(|B^2_0|^2+|B^2_1|^2)(|C_0|^2+2|C_1|^2),
\nonumber\\
&&M(22220)\propto
-\frac{2}{175}
[3|A^2_{01}|^2+2\sqrt{6}Re(A^2_{12}A^{2\ast}_{10})]
(2|B^2_0|^2-|B^2_1|^2)(|C_0|^2-|C_1|^2),
\nonumber\\
&&M(22221)\propto
-\frac{6}{175}
[\sqrt{3}|A^2_{01}|^2+2\sqrt{2}Re(A^2_{12}A^{2\ast}_{10})]
Re(B^2_1B^{2\ast}_0)](|C_0|^2-|C_1|^2), 
\nonumber\\
&&M(22222)\propto
-\frac{6}{175}
[3|A^2_{01}|^2+2\sqrt{6}Re(A^2_{12}A^{2\ast}_{10})]
|B^2_1|^2(|C_0|^2-|C_1|^2),
\nonumber\\   
&&M(23121)\propto
\frac{6}{175}[2Im(A^2_{01}A^{2\ast}_{00}+A^2_{10}A^{2\ast}_{11})+
\sqrt{6}Im(A^2_{11}A^{2\ast}_{12})]
Im(B^2_1B^{2\ast}_0)(|C_0|^2-|C_1|^2),
\nonumber\\
&&M(23221)\propto
-\frac{12\sqrt{5}}{175}
Im(A^2_{12}A^{2\ast}_{10})Im(B^2_1B^{2\ast}_0)(|C_0|^2-|C_1|^2),
\nonumber\\
&&M(24000)\propto
-\frac{2}{315}
(6|A^2_{00}|^2+4|A^2_{01}|^2-6|A^2_{10}|^2-8|A^2_{11}|^2-|A^2_{12}|^2)
(3|B^2_0|^2-4|B^2_1|^2)(|C_0|^2+2|C_1|^2),
\nonumber\\
&&M(24020)\propto
-\frac{4}{1575}
(6|A^2_{00}|^2+4|A^2_{01}|^2-6|A^2_{10}|^2-8|A^2_{11}|^2-|A^2_{12}|^2)
(3|B^2_0|^2+2|B^2_1|^2)(|C_0|^2-|C_1|^2),
\nonumber\\
&&M(24021)\propto
-\frac{2\sqrt{10}}{525}
(6|A^2_{00}|^2+4|A^2_{01}|^2-6|A^2_{10}|^2-8|A^2_{11}|^2-|A^2_{12}|^2)
Re(B^2_1B^{2\ast}_0)(|C_0|^2-|C_1|^2),
\nonumber\\
&&M(24022)\propto
-\frac{4\sqrt{15}}{1575}
(6|A^2_{00}|^2+4|A^2_{01}|^2-6|A^2_{10}|^2-8|A^2_{11}|^2-|A^2_{12}|^2)
|B^2_1|^2(|C_0|^2-|C_1|^2),
\nonumber\\
&&M(24100)\propto
-\frac{\sqrt{10}}{315}
[6Re(A^2_{11}A^{2\ast}_{10}-A^2_{01}A^{2\ast}_{00})+\sqrt{6}
Re(A^2_{12}A^{2\ast}_{11})]
(3|B^2_0|^2-4|B^2_1|^2)(|C_0|^2+2|C_1|^2),
\nonumber\\   
&&M(24120)\propto
-\frac{2\sqrt{10}}{1575}
[6Re(A^2_{11}A^{2\ast}_{10}-A^2_{01}A^{2\ast}_{00})
+\sqrt{6}Re(A^2_{12}A^{2\ast}_{11})](3|B^2_0|^2+2|B^2_1|^2)(|C_0|^2-|C_1|^2),
\nonumber\\
&&M(24121)\propto
-\frac{2}{105}[6Re(A^2_{01}A^{2\ast}_{00}-A^2_{11}A^{2\ast}_{10})
-\sqrt{6}Re(A^2_{12}A^{2\ast}_{11})]Re(B^2_1B^{2\ast}_0)(|C_0|^2-|C_1|^2),
\nonumber\\
&&M(24122)\propto
-\frac{4}{105}[\sqrt{6}Re(A^2_{01}A^{2\ast}_{00}-A^2_{11}A^{2\ast}_{10})
-Re(A^2_{12}A^{2\ast}_{11})]|B^2_1|^2(|C_0|^2-|C_1|^2),
\nonumber\\
&&M(24200)\propto
-\frac{2\sqrt{10}}{315}[\sqrt{6}|A^2_{01}|^2
-3Re(A^2_{12}A^{2\ast}_{10})](3|B^2_0|^2-4|B^2_1|^2)(|C_0|^2+2|C_1|^2),
\nonumber\\
&&M(24220)\propto
-\frac{4\sqrt{10}}{1575}[\sqrt{6}|A^2_{01}|^2
-3Re(A^2_{12}A^{2\ast}_{10})](3|B^2_0|^2+2|B^2_1|^2)(|C_0|^2-|C_1|^2),
\nonumber\\
&&M(24221)\propto
\frac{4}{105}[-\sqrt{6}|A^2_{01}|^2
+3Re(A^2_{12}A^{2\ast}_{10})]Re(B^2_1B^{2\ast}_0)(|C_0|^2-|C_1|^2),
\nonumber\\
&&M(24222)\propto
-\frac{4}{105}[2|A^2_{01}|^2
-\sqrt{6}Re(A^2_{12}A^{2\ast}_{10})]|B^2_1|^2(|C_0|^2-|C_1|^2).
\nonumber
\end{eqnarray}

\end{document}